\documentclass{emulateapj}          

\def\iso#1#2{\mbox{${}^{#2}{\rm #1}$}}

\def\li#1{\iso{Li}{#1}}
\def\c1#1{\iso{C}{1#1}}
\def\n1#1{\iso{N}{1#1}}
\def\o1#1{\iso{O}{1#1}}

\def\beq{\begin{equation}}
\def\eeq{\end{equation}}
\def\beqar{\begin{eqnarray}}
\def\eeqar{\end{eqnarray}}

\def\la{\mathrel{\mathpalette\fun <}}
\def\ga{\mathrel{\mathpalette\fun >}}
\def\fun#1#2{\lower3.6pt\vbox{\baselineskip0pt\lineskip.9pt
  \ialign{$\mathsurround=0pt#1\hfil##\hfil$\crcr#2\crcr\sim\crcr}}}

\begin{document}

\title{Can Galactic Cosmic Rays Account for Solar \li6 \\
Without Overproducing Gamma Rays?}

\author{Tijana Prodanovi\'{c} and Brian D. Fields} 
 
\affil{Center for Theoretical Astrophysics,
Department of Astronomy, University of Illinois,
Urbana, IL 61801}

\begin{abstract}

Cosmic-ray interactions with interstellar gas produces both \li6, which 
accumulates in the interstellar medium (ISM), and $\pi^0$ mesons, which decay 
to gamma-rays which propagate throughout the cosmos. Local \li6 abundances and
extragalactic gamma-rays thus have a common origin which tightly links them. 
We exploit this connection to use gamma-ray observations to infer the 
contribution to \li6 nucleosynthesis by standard Galactic cosmic-ray (GCR) 
interactions with the ISM. Our calculation uses a carefully propagated 
cosmic-ray spectrum and accounts for \li6 production from both fusion 
reactions ($\alpha \alpha \rightarrow \li6$) as well as from spallation 
channels (${\rm p},\alpha+\rm CNO \rightarrow \li6$). We find that although 
extreme assumptions yield a consistent picture, more realistic ones indicate 
that solar \li6 cannot be produced by standard GCRs alone without 
overproducing the hadronic gamma rays. Implications for the primordial \li6 
production by decaying dark matter and cosmic rays from cosmological structure
formation are discussed. Upcoming gamma-ray observations by {\it GLAST} will 
be crucial for determining the resolution of this problem. 

\end{abstract}


\section{Introduction}

Cosmic-ray nucleosynthesis is the only known Galactic source of the \li6
\citep{elisa,fo99}. Thus, it is a standard belief that the observed solar 
abundance of \li6 was produced by Galactic cosmic-ray (GCR) 
interactions with the interstellar medium (ISM), where $\alpha \alpha 
\rightarrow \li6$ is the dominant channel \citep{sw,montmerle}. However, 
hadronic CRs also produce gamma rays, and thus GCR interactions in 
normal galaxies are guaranteed to contribute \citep{pavlidou} to the 
observed extragalactic gamma-ray background \citep[hereafter EGRB;][]{strong}.
Moreover, since they both originate from CR interactions,
\li6 and hadronic gamma-rays are tightly related.

\citet{fp} established a simple and model-independent
connection between lithium and ``pionic'' $\gamma$-ray 
production ($pp \rightarrow \pi^0 \rightarrow \gamma \gamma $)
by a given cosmic-ray population. Using this tool with
simplifying assumptions gave the alarming result
that the solar \li6 abundance, if produced entirely by GCRs, 
demands a pionic $\gamma$-ray intensity exceeding
the {\it entire} observed EGRB \citep{fp}. 
Given the current interest in \li6,
this result thus deserves a thorough investigation.

In this paper we revisit the problem of Li--$\gamma$-ray consistency
with a more precise and realistic calculation. We now employ a carefully 
propagated CR spectrum, as opposed to the standard single-power law 
spectrum adopted in \citet{fp}. Moreover, instead of using a convenient fit 
for the pionic $\gamma$-ray spectrum \citep{ensslin} we now calculate it 
self-consistently from our CR spectrum.  We also estimate the spallation
 ${\rm p},\alpha + \rm CNO \rightarrow \li6$ contribution to the solar \li6 
abundance. These effects slightly reduce but do not eliminate the discrepancy.
Moreover, the only remaining effect we expect to be important--\li6 
destruction as it is processed through stars--makes the problem more severe.
The net effect is that in a realistic calculation, 
the observed EGRB allows for only $\approx 60\% \ \li6_\odot$  
to be produced by standard GCRs. Only a conspiracy of extreme assumptions gives
GCR production of the solar \li6 that does not at the same time saturate 
the observed EGRB.

Our result represents a strong hint for the need of a new \li6 source. 
Recent suggestions such as dark matter and low-energy cosmic rays are 
discussed in \S \ref{sect:discussion}. Upcoming gamma-ray observations by 
{\it GLAST} \citep{GLAST} will better constrain (or determine!) the pionic 
$\gamma$-ray fraction of the EGRB and will thus be the key in determining the 
severity of this problem.

\section{Lithium--Gamma-Ray Connection}

In \citet{fp} we formally demonstrated and quantified the tight
connection between CR lithium synthesis and hadronic $\gamma$-ray production.
We showed that both observables are a measure of the time-integrated
CR flux (fluence $F$). Specifically, the ratio of the ``pionic'' 
$\gamma$-ray intensity $I_{\gamma_\pi}$ (integrated over the entire 
energy spectrum) 
and \li6 abundance (baryon or mole fraction $\li6 \equiv Y_6 \equiv n_6/n_{\rm baryon}$) produced in fusion reactions
with the ISM can be expressed essentially as the ratio of their reaction rates 
\beq
\label{eq:gamma2li}
\frac{I_{\gamma_\pi}(E>0,t)}{\li6(\vec{x},t)}
 = \frac{ n_{\rm b} c}{4\pi y_{\alpha ,\rm  cr} y_{\alpha ,\rm ism}}    \
   \frac{ \sigma_{\gamma}}{\sigma_6} \
   \frac{F_{\rm avg}(t)}{F_{\rm MW}(t)}   
\eeq
This factorizes into a product of nuclear, cosmological, and cosmic-ray
parameters, and a ``Copernican'' factor $F_{\rm avg}/F_{\rm MW}$.
Cosmology enters via the
comoving baryon number density
$n_{\rm b}=2.52 \times 10^{-7} \ \rm cm^{-3}$.
The cosmic-ray and ISM helium abundances are taken to be 
$y_{\rm \alpha}^{\rm cr}=y_{\rm \alpha}^{\rm ism}=0.1$ ($y_i \equiv n_i/n_H$).
The flux-averaged pionic $\gamma$-ray 
production cross-section is $\sigma_{\gamma} \equiv 2 \xi_\alpha \zeta_\pi 
\sigma_{\pi^0} $ where the factor of 2 counts the number of photons per pion 
decay, $\sigma_{\pi^0}$ is the cross section, $\zeta_\pi$ is the pion 
multiplicity, and the factor $\xi_\alpha  = 1.45$ accounts for 
$p\alpha$ and $\alpha\alpha$ reactions \citep{dermer}. 
For $\sigma_6$ we have used a recent result of
\citet{mercer}, which for \li6 differs significantly from the old
values. The use of the new cross sections results in a lower \li6 production. 

The ratio $F_{\rm avg}/F_{\rm MW}$ is the ratio of the line-of-sight 
baryon-averaged fluence to the local fluence; this ``Copernican
ratio'' compares the cumulative cosmic-ray activity of our Galaxy
to that of an average star-forming galaxy.
Following our previous
work we will initially take this factor be $\approx 1$, i.e. 
that the Milky Way CR flux through out the history can be
approximated with the cosmic mean.
We will then examine the consequences that this ratio differs
significantly from unity.

\section{Cosmic-Ray Spectrum}

In eq.~(\ref{eq:gamma2li}), the Li-$\gamma$-ray proportionality
depends on the ratio of the mean cross sections 
$ \sigma_{\gamma}/\sigma_6$. These must be properly
averaged over the GCR energy spectrum. In \citet{fp} we have adopted a 
standard propagated cosmic-ray spectrum which is a single power-law in total 
energy with a spectral index $\alpha=2.75$ over the entire relevant energy 
range. 
While this is a commonly-used rough approximation to the
GCR spectrum, it becomes inaccurate at energies $\la 1$ GeV,
where ionization energy losses dominate over escape losses.
Because $\alpha \alpha \rightarrow \li6$
threshold energy is at $\sim 10$ MeV/nucleon, while $\rm pp \rightarrow \pi^0$
threshold is at $\sim 280$ MeV, the Li-$\gamma$ connection is particularly 
sensitive to GCR behavior at very low energies. Thus in this paper we refine 
on the analysis presented in \citet{fp} by calculating and implementing a 
carefully propagated CR spectrum for a leaky box model \citep{meneguzzi}.

The resulting CR spectrum calculated using the standard leaky box model,
assuming a standard source spectrum that is a power-law in
momentum \citep[e.g.,]{gaisser}.
This gives a CR flux
$\sim 4$ times higher around $\alpha \alpha \rightarrow \li6$
threshold, compared to the one used in \citet{fp} where a single power-law
spectrum was assumed, while for energies $\ga 100$ MeV a single-power law 
spectrum is a good approximation.

\section{Pionic Gamma-ray Spectrum}

\begin{figure}
\epsscale{.80}
\plotone{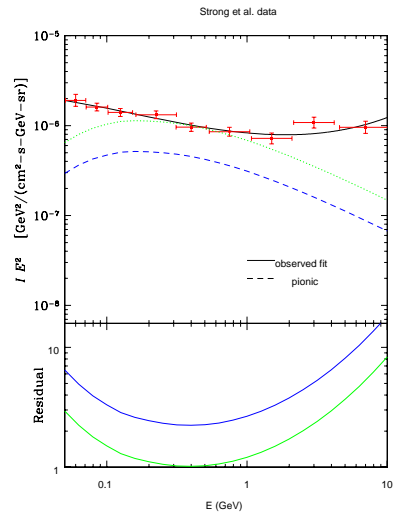}
\caption{In the upper panel of this figure, we plot the pionic spectrum 
(dotted green line - maximized, dashed blue line - normalized to the Milky Way)
, compared to the observed EGRB spectrum (solid line, fit to data); 
we use the \citet{strong} data points, which are given in red crosses.
The bottom panel represents the residual function, that is, 
$\log [(I E^2)_{\rm obs}/(I E^2)_{\pi}]=\log (I_{\rm obs}/I_{\pi})$.}
\end{figure}

\citet{ensslin} provide a useful parametrization of the pionic $\gamma$-ray
spectrum used in \citet{fp}. However, here we numerically calculate the pionic
$\gamma$-ray spectrum in full detail, by adopting the isobar+scaling model as
given in \cite{dermer}; the pionic spectrum we adopt uses
the same cosmic-ray spectrum as the \li6 production, and 
thus is self-consistent.
 
In order to calculate
$I_{\gamma_\pi}$ one needs to know the history of the CR sources and the 
targets. Both histories come from the cosmic star-formation rate.
As described in detail in \citet{fp} we can obtain the GCR pionic 
$\gamma$-ray spectrum integrated over the history of the sources 
\citep[equation 26 of][same parameter values used]{fp}.
The cosmic star-formation rate alone fixes the {\em shape}
of the pionic EGRB, but requires a normalization that
physically connects the star formation rate to the cosmic-ray flux,
and which normalizes the present gas fraction in a typical galaxy.
In order to place an {\em upper limit} to the pionic EGRB,
we allow this normalization to vary freely to maximize 
the pionic $\gamma$-ray flux consistent with
present EGRB observations \citep{fp,pf}. This is presented in Fig. 1
as a dotted green line. The observed EGRB spectrum is that of \citet{strong}
and is plotted as red data points, with a black solid line fit \citep{fp}.
Finally, we find maximal pionic $\gamma$-ray fraction to be $ 58 \%$ of
the total observed EGRB.

More realistically, we can use the Milky Way to determine
both the scaling between the star formation rate and the cosmic-ray flux,
and the present-day gas fraction.
We do this following \citet{pavlidou},
using a 
present gas content of the Milky Way $M_{\rm gas,MW} \approx 10^{10}
M_{\odot}$, and a star formation rate $3.2 \ M_\odot/{\rm yr}$.
The resulting $\gamma$-ray spectrum is presented in Fig. 1 as a blue dashed 
line. This corresponds to the pionic $\gamma$-ray contribution expected from 
the normal galaxies. In addition to this guaranteed component to the EGRB, 
unresolved blazars will also contribute significantly \citep{stecker,pavlidou},
presumably comprising much or all of the remaining signal.

Having determined a upper limit and a more realistic
estimate to $I_{\gamma_\pi}$ one can find the
corresponding Li abundance, 
via eq.~(\ref{eq:gamma2li}).
This is our main goal, to which we now turn.

\section{Upper Limits on and Estimates of GCR-Produced \li6}

In this section we calculate limits to and estimates of
the \li6 produced by GCRs 
that are allowed by preset EGRB data. 
We present our results in steps of 
increasing realism.  For now we retain the 
Copernican assumption that the Milky Way cosmic-ray
fluence is typical of star-forming galaxies ($F_{\rm MW}/F_{\rm MW} = 1$);
we will revisit this assumption 
in the final section. 

1.
By combining (\ref{eq:gamma2li}) with the {\it maximal} pionic $\gamma$-ray fraction and 
procedure described in \citet{fp}, we find the fraction of \li6 abundance
produced in $\alpha \alpha \rightarrow \li6 $ reaction to be
$\rm \li6_{\alpha \alpha}=0.61 \ \li6_{\odot}$
\citep[$(\li6/{\rm H})_\odot=1.53 \times 10^{-10}$;][]{ag}.
This corresponds to an extreme upper limit for all \li6 produced by
the GCR $\alpha \alpha$ reaction.

2.
Though the $\alpha \alpha$  reaction with the ISM is the dominant channel 
for \li6 production, a non-negligible contribution, especially at higher 
metallicity, comes from the spallation reactions $\rm p,\alpha + CNO \rightarrow
\li6 $ (both forward and inverse kinematics, that is fast heavy nuclei, are
included). If the fusion and CNO reaction rates were to be equal
the required oxygen abundance should be $\rm (O/H)_{eq}=0.51\ (O/H)_{\odot}$.
This now sets the normalization and allows us to calculate the total
\li6 abundance produced from all channels under the extreme assumption
that the ISM was at {\it solar metallicity} over the Galactic history.
We find that $\rm \li6_{GCR}=1.79 \ \li6_{\odot}$, which now represents the 
extreme upper limit for {\it all} \li6 produced by GCRs.

3.
Because the cosmic-ray CNO abundance is a direct function of the Galactic
supernova rate, a precise calculation introduces a factor of 1/2, that
is, instead of assuming solar metallicity through out history one should
use an average value of $\rm (O/H)_{eq}=0.5\ (O/H)_{\odot}$.
This results in the total allowed GCR-produced \li6 abundance of
$\rm \li6_{GCR}=1.20 \ \li6_{\odot}$, which is still consistent with the 
standard picture.

4.
So far we have been taking the maximal (Fig. 1, dotted green line) 
pionic $\gamma$-ray fraction as allowed by the present EGRB data
\footnote{Determination of the EGRB relies on the subtraction of the Galactic 
Plane and is thus model-dependent. Although \citet{kwl} report only a limit 
to the EGRB this would only strengthen our result.},
where we have (without justification) ignored the normalization and
just used the shape of our spectrum. However, it is unrealistic to assume that the
entire emission is due to GCRs. Indeed, independent of the details of our 
galactic $\gamma$-ray estimate, it is clear that the EGRB must contain a large
and perhaps dominant contribution from unresolved AGNs (blazars) and so 
the galactic signal must leave room for this and cannot saturate the observed 
level. An estimate of the normalized GCR pionic $\gamma$-ray component of the 
EGRB (Fig. 1, dashed blue line) yields a spectrum that is a factor of 2.1 
lower than the maximized value. Thus, in this most honest case, we find
$\rm \li6_{GCR}=0.57 \ \li6_{\odot}$ which now falls short by about a factor 
of 2 from a standard picture of cosmic-ray \li6 nucleosynthesis.

5.
For inverse CNO kinematics a non-negligible LiBeB production comes from
two-step spallation reactions, eg. $\rm O+H \rightarrow ^{11}B+H \rightarrow
\li6$ \citep{kneller03}. For example, the production rate of \li6  from 
two-step reactions of fast oxygen is $\sim 40\%$ of single-step fast oxygen
spallation reactions, for a fixed $\Lambda=10 \rm \ g/cm^2$ \citep{kneller06}.
However, when two-step inverse CNO kinematics is taken into account, the
overall increase is only slight and the result now becomes
\beq
\label{eq:result}
\rm \li6_{GCR}=0.59 \ \li6_{\odot}
\eeq
Even in the most extreme assumption that the two-step rates are equal to the
single-step inverse CNO kinematic rates, the resulting \li6 abundance
would still be only $63\%$ of the solar.

6.
Finally, one has to remember that the observed solar \li6 abundance is not
the total lithium abundance produced, due to astration, that is, the fact that
some of the gas was already processed by stars. Due to very fragile nature of
this isotope, $\li6_{\odot}$ is only the lower bound on the
total \li6 produced. For a rough estimate of the level of astration one can 
use deuterium. It is well established that Big Bang nucleosynthesis is
the only important source of D \citep{els,pf03} and that D is easily destroyed
in stars due to a similarly fragile nature. Thus by comparing the solar nebula
D abundance $\rm D_{presol}=2.1 \times 10^{-5}$ \citep{gg} with
the abundance determined from 5 best quasar absorption systems 
$\rm D_{QSO}=2.78 \times 10^{-5}$ \citep{cfo03}, we find that
roughly $\sim 25\%$ of the gas has passed through stars.
Thus $\li6_\odot$ is about $\sim 75\%$ of $\li6_{\rm tot}$, and our calculated 
GCR \li6 now becomes $\rm \li6_{GCR} \sim 0.45 \ \li6_{tot}$.

\section{Discussion}
\label{sect:discussion}

In this paper we have used the connection between \li6 and
pionic $\gamma$-rays produced in CR interactions, in order to calculate
the total allowed \li6 abundance that can be produced by GCRs. 
We have used a CR spectrum, carefully propagated according to the
leaky-box model, while the pionic $\gamma$-ray spectrum was calculated
based on the \citet{dermer} model. A realistic, detailed calculation that
includes \li6 production from both fusion reaction with the ISM and spallation
CNO channels (2-step inverse kinematics also included), yields a \li6 
abundance that is only $\approx 60\%$ of the total \li6 produced, if standard 
GCRs are the only relevant source. Correcting for astration will
result in even lower $\li6_{\rm GCR}$ abundance at the level of 
$\sim 45\% \ \li6_{\rm tot}$.

Our result either indicates the need for a new important source of \li6 
beyond standard GCR nucleosynthesis, or it points to a possible failure 
of the usual assumption that the average interstellar GCR
flux tracks the instantaneous star formation rate. We consider each 
possibility in turn.

Additional sources of \li6 are of considerable current interest,
because of the recent report of a \li6 plateau in metal-poor halo stars
\citep{asplund}. As with the familiar \li7 Spite plateau, an analogous 
\li6 feature would suggest a pre-Galactic source of \li6. 
And indeed, recently two very different cosmological sources of 
\li6 have been proposed:
(1) production in the early universe, stimulated by supersymmetric dark 
matter particle decays during big bang nucleosynthesis 
\citep{dimopoulos,kkm,karsten,eov,kusakabe};
and (2) production during the virialization and baryonic accretion
of large-scale structures, which generates cosmological shocks \citep{miniati}
that can in turn accelerate a population of cosmological cosmic rays 
\citep[][but see \cite{prantzos} for constraints]{suzuki,blasi}.

The \li6 plateau is $\la 10\% \ \li6_{\odot}$,
and thus whatever its source is, it will not be able to account for the
factor $\ga 2$ discrepancy between $\li6_{\odot}$ and 
$\li6_{\rm GCR}$ we have found. However, the existence 
of the \li6 plateau at the 10\% level of the solar abundance for metallicities
$\rm [Fe/H] \la -1$, can be used as a constraint to any non-standard \li6 
source that is expected to account for the potentially missing $\approx 40\%$ 
of $\li6_{\odot}$. Moreover, \li6 plateau would indicate that such a source 
would have to become important only at late times, and near-solar 
metallicities.

We note that another additional source of \li6 could come from a population 
of CRs having low energies ($\la 100$ MeV). Such particles are 
excluded from the solar system and hence not directly constrained 
observationally. A large flux of such particles, well above the extrapolated
observed high-energy trends, could produce large additional amounts of \li6 
but no pions and hence no pionic $\gamma$-rays. Indeed, recent observations 
of ${\rm H}_3^{+}$ in molecular clouds \citep{mccall} seem to demand a large 
low-energy CR flux in the neighborhood of these clouds. On the other 
hand, low-energy CRs widespread enough to participate significantly 
in LiBeB nucleosynthesis on Galactic scales face strong constraints that come 
from energetics \citep{ramaty} and from LiBeB abundance ratios \citep{elisa98}.
These limits are evaded if solar \li6 reflects a localized low-energy 
CR enhancement, either due to a hypernova-like Type Ic supernova
\citep{fdcv,ns}, or to solar CR production in the protosolar
nebula \citep[e.g.,][]{gounelle}.  In either case, the other \li7BeB isotopes will
be produced and constrain the allowable \li6 contribution.

In this work we have assumed that the Milky Way CR fluence can be approximated
with the cosmic mean. Therefore, our result might indicate that more \li6
was produced than $\gamma$-rays would suggest, which would be the case
if the Milky Way CR flux was at some time(s) a factor of $\sim 2$ 
(on average) higher than the typical CR flux in a normal galaxy. 

If indeed \li6 points to enhanced CR activity, this in turn would point to
anomalies in Milky Way star formation and/or CR properties.
We have assumed that the CR fluence
here is typical of the mean star-forming galaxy ($F_{\rm MW}/F_{\rm avg} = 1$).
If instead our Galaxy had a more vigorous CR 
history, this could account for the difference.
Also, we have in our most realistic assessment used the
present {\em local} cosmic-ray/star-formation ratio 
$\Phi_{\rm MW}/\psi_{\rm MW}$; if this departs from the cosmic mean,
this too could account for the \li6 discrepancy.
Both of these solutions have implications for
(and quantify) the utility of the Milky Way and Local Group
as representative cosmological samples of star forming galaxies.

Upcoming gamma-ray experiments will go far
to clarify the nature of Galactic and extragalactic pionic
gamma-rays, and hence \li6 production. 
{\it GLAST} could detect 
the pionic $\gamma$-ray signature from
diffuse Galactic emission as well as in the EGRB;
this would remove the need to estimate these components.
{\it GLAST} should also detect several Local Group
galaxies and thus allow for new determinations
of the cosmic-ray/star-formation scaling \citep{vaso1}.
Also, our calculation is hampered by lack of 
evidence of the ``pion bump'' in the Milky Way $\gamma$-ray spectrum.
Fortunately, it was
recently demonstrated that future GeV--TeV--PeV gamma-ray observations of 
the diffuse emission from the Galactic Plane can
determine the level of pionic $\gamma$-ray emission in the Milky
Way \citep{pfb}. 

In closing, we have underscored the increasing crucial role
that \li6 plays in particle astrophysics and cosmology;
the excess we find in solar \li6 has implications
throughout these fields. Fortunately, 
upcoming measurements of \li6 and gamma-rays 
should be able to address the questions posed here.

\acknowledgments
We are grateful to Charles Dermer for illuminating discussions, and to
James Kneller for valuable information.
We thank Zarija Luki\'{c} for help with coding.
This material is based on work supported by the National Science
Foundation under grant AST-0092939.

{}


\begin{thebibliography}{}

\bibitem[Anders \& Grevesse(1989)]{ag} Anders, E.~\&
Grevesse, N.\ 1989, \gca, 53, 197

\bibitem[Asplund et al.(2005)]{asplund} Asplund, M., Lambert, 
D.~L., Nissen, P.~E., Primas, F., \& Smith, V.~V.\ 2005, ApJ {\it in press}, 
arXiv:astro-ph/0510636 

\bibitem[Blasi(2004)]{blasi} Blasi, P.\ 2004, Astroparticle 
Physics, 21, 45 

\bibitem[Cyburt et al.(2003)]{cfo03} Cyburt, R.~H., Fields, 
B.~D., \& Olive, K.~A.\ 2003, Physics Letters B, 567, 227 

\bibitem[Dermer(1986)]{dermer} Dermer, C.~D.\ 1986, \aap, 157, 
223 

\bibitem[Dimopoulos et al.(1988)]{dimopoulos} Dimopoulos, S., 
Esmailzadeh, R., Starkman, G.~D., \& Hall, L.~J.\ 1988, \apj, 330, 545 

\bibitem[Ellis et al.(2005)]{eov} Ellis, J., Olive, K.~A., 
\& Vangioni, E.\ 2005, Physics Letters B, 619, 30 

\bibitem[Epstein et al.(1976)]{els} Epstein, R.~I., 
Lattimer, J.~M., \& Schramm, D.~N.\ 1976, \nat, 263, 198 

\bibitem[Fields et al.(2002)]{fdcv} Fields, B.~D., Daigne, 
F., Cass{\'e}, M., \& Vangioni-Flam, E.\ 2002, \apj, 581, 389 

\bibitem[Fields \& Olive(1999)]{fo99} 
Fields, B.~D.~\& Olive, K.~A.\ 1999b, New Astronomy, 4, 255

\bibitem[Fields \& Prodanovi{\'c}(2005)]{fp} Fields, 
B.~D., \& Prodanovi{\'c}, T.\ 2005, \apj, 623, 877

\bibitem[Geiss \& Gloeckler(1998)]{gg} Geiss, J., \& 
Gloeckler, G.\ 1998, Space Science Reviews, 84, 239 

\bibitem[Gaisser(1990)]{gaisser} Gaisser, T.~K.\ 1990, 
Cambridge and New York, Cambridge University Press, 1990, 292 p.,  

\bibitem[Gehrels \& Michelson(1999)]{GLAST} Gehrels, N., \& 
Michelson, P.\ 1999, Astroparticle Physics, 11, 277

\bibitem[Gounelle et al.(2006)]{gounelle} Gounelle, M., Shu, 
F.~H., Shang, H., Glassgold, A.~E., Rehm, K.~E., \& Lee, T.\ 2006, \apj, 
640, 1163 

\bibitem[Jedamzik et al.(2005)]{karsten} Jedamzik, K., Choi, 
K.-Y., Roszkowski, L., \& Ruiz de Austri, R.\ 2005,
{\tt arXiv:hep-ph/0512044}

\bibitem[Kawasaki et al.(2005)]{kkm} Kawasaki, M., Kohri, 
K., \& Moroi, T.\ 2005, Physics Letters B, 625, 7 

\bibitem[Keshet et al.(2004)]{kwl} Keshet, U., Waxman, E., 
\& Loeb, A.\ 2004, Journal of Cosmology and Astro-Particle Physics, 4, 6 

\bibitem[Kneller et al.(2003)]{kneller03} Kneller, J.~P., 
Phillips, J.~R., \& Walker, T.~P.\ 2003, \apj, 589, 217 

\bibitem[Kneller(2006)]{kneller06}
Kneller, J.~P. \ 2006, private communication

\bibitem[Kusakabe et al.(2006)]{kusakabe} Kusakabe, M., Kajino, 
T., \& Mathews, G.~J.\ 2006,  
{\tt arXiv:astro-ph/0605255} 

\bibitem[McCall et al.(2003)]{mccall} McCall, B.~J., et al.\ 
2003, \nat, 422, 500 

\bibitem[Meneguzzi et al.(1971)]{meneguzzi} Meneguzzi, M., 
Audouze, J., \& Reeves, H.\ 1971, \aap, 15, 337 

\bibitem[Mercer et al.(2001)]{mercer} Mercer, D.~J., et al.\ 
2001, \prc, 63, 065805 

\bibitem[Miniati et al.(2000)]{miniati} Miniati, F., Ryu, D., 
Kang, H., Jones, T.~W., Cen, R., \& Ostriker, J.~P.\ 2000, \apj, 542, 608 

\bibitem[Montmerle(1977)]{montmerle} 
Montmerle, T.\ 1977, \apj, 217, 878 % CCR & lite elements

\bibitem[Nakamura \& Shigeyama(2004)]{ns} Nakamura, K., \& 
Shigeyama, T.\ 2004, \apj, 610, 888

\bibitem[Pavlidou \& Fields(2001)]{vaso1}
Pavlidou, V., \& 
Fields, B.~D.\ 2001, \apj, 558, 63 

\bibitem[Pavlidou \& Fields(2002)]{pavlidou} 
Pavlidou, V.~\& Fields, B.~D.\ 2002, \apjl, 575, L5 

\bibitem[Pfrommer \& En{\ss}lin(2004)]{ensslin}
Pfrommer, C.~\& En{\ss}lin, T.~A.\ 2004, \aap, 413, 17

\bibitem[Prantzos(2006)]{prantzos} 
Prantzos, N.\ 2006, \aap, 448, 665 

\bibitem[Prodanovi{\'c} \& Fields(2004)]{pf} 
Prodanovi{\'c}, T., \& Fields, B.~D.\ 2004, Astroparticle Physics, 21, 627 

\bibitem[Prodanovi{\'c} \& Fields(2003)]{pf03} 
Prodanovi{\'c}, T., \& Fields, B.~D.\ 2003, \apj, 597, 48 

\bibitem[Prodanovic et al.(2006)]{pfb} Prodanovic, T., 
Fields, B.~D., \& Beacom, J.~F.\ 2006, ArXiv Astrophysics e-prints, 
arXiv:astro-ph/0603618 

\bibitem[Ramaty \& Lingenfelter(1999)]{ramaty} Ramaty, R., \& 
Lingenfelter, R.~E.\ 1999, ASP Conf.~Ser.~171: LiBeB Cosmic Rays, and 
Related X- and Gamma-Rays, 171, 104 

\bibitem[Stecker \& Salamon(1996)]{stecker} Stecker, F.~W., \& 
Salamon, M.~H.\ 1996, \apj, 464, 600 

\bibitem[Steigman \& Walker(1992)]{sw} 
Steigman, G.~\& Walker, T.~P.\ 1992, \apjl, 385, L13 

\bibitem[Strong et al.(2004)]{strong} Strong, A.~W., 
Moskalenko, I.~V., \& Reimer, O.\ 2004, \apj, 613, 956 

\bibitem[Suzuki \& Inoue(2002)]{suzuki} Suzuki, T.~K., \& 
Inoue, S.\ 2002, \apj, 573, 168  

\bibitem[Vangioni-Flam et al.(1998)]{elisa98} Vangioni-Flam, 
E., Ramaty, R., Olive, K.~A., \& Casse, M.\ 1998, \aap, 337, 714 

\bibitem[Vangioni-Flam et al.(1999)]{elisa} 
Vangioni-Flam, E., Casse, M., Cayrel, R., Audouze, J., Spite, M., 
\& Spite, F.\ 1999, New Astronomy, 4, 245 

\end{thebibliography}
\end{document}